\documentclass[iop,12pt]{emulateapj}
\usepackage{natbib} 
\usepackage{graphicx}
\usepackage{amssymb,amsmath}
\usepackage[dvipsnames]{color}
\newcommand{\Red}[1]{\textcolor{red}{#1}}

\def\Kepler{{\it Kepler}\,\,}

\def\refnew#1{\,(\ref{#1})}
\def\Teff{T_{\mathrm{eff}}}
\def\Msun{M_\odot}

\def\zw{z_\omega}
\def\zb{z_b}

\def\Lamb{L_{\ell}}
\def\Lambb{L_{\ell,b}}
\def\Del{\Delta}

\def\K{\,\mathrm{K}}
\def\min{\,\mathrm{min}}

\def\erg{\,\mathrm{erg}}

\newcommand{\appropto}{\mathrel{\vcenter{
  \offinterlineskip\halign{\hfil$##$\cr
    \propto\cr\noalign{\kern2pt}\sim\cr\noalign{\kern-2pt}}}}}

\begin{document}

\title{DAVs: Red edge and Outbursts}
\author{Jing Luan}
\email{jingluan@berkeley.edu}
\affiliation{University of California at Berkeley, Berkeley, CA 94720, US}
\author{Peter Goldreich}
\affiliation{California Institute of Technology, Pasadena, CA 91125, US}

\begin{abstract}
As established by photometric surveys, white dwarfs with hydrogen atmospheres and surface gravity, $\log(g)\approx 8.0$ pulsate as they cool across the temperature range $12500\,\mathrm{K} \gtrsim T_{\mathrm{eff}} \gtrsim 10800\,\mathrm{K}$. Known as DAVs or ZZ Ceti stars, their oscillations are attributed to gravity modes excited by convective driving. Overstability requires convective driving to exceed radiative damping. Previous works have demonstrated that $\omega\gtrsim\max( \tau_c^{-1},L_{\ell,b})$ is a necessary and sufficient condition for overstability. Here $\tau_c$ and $L_{\ell,b}$ are the effective thermal timescale and Lamb frequency at the base of the surface convection zone. Below the observational red edge, $L_{\ell,b}\gg \tau_c^{-1}$,  so overstable modes all have $\omega\tau_c\gg 1$. Consequently, their photometric amplitudes are reduced by that large factor rendering them difficult to detect. Although proposed previously, the condition $\omega\gtrsim L_{\ell,b}$ has not been clearly interpreted. We show that modes with $\omega<L_{\ell,b}$ suffer enhanced radiative damping that exceeds convective driving rendering them damped. A quasi-adiabatic analysis is adequate to account for this enhancement. Although this approximation is only marginally valid at the red edge, it becomes increasingly accurate towards both higher and lower $\Teff$. Recently, {\it Kepler} discovered a number of cool DAVs that exhibit sporadic flux outbursts. Typical outbursts last several hours, are separated by days, and release $\sim 10^{33}-10^{34}\,\mathrm{erg}$.  We attribute outbursts to limit cycles arising from sufficiently resonant 3-mode couplings between overstable parent modes and pairs of radiatively damped daughter modes. Limit cycles account for the durations and energies of outbursts and their prevalence near the red edge of the DAV instability strip.
\bigskip
\end{abstract}

\maketitle
\section{Introduction}\label{sec:intro}
DAVs exhibit multi-mode oscillations.\footnote{Photometry dominates the discovery and investigation of DAVs.} The longest period modes have $P\sim 2\,\mathrm{min}$ near the blue edge and $P\sim 20\,\mathrm{min}$ near the red edge of the DAV instability strip. Longer-period modes in cooler DAVs exhibit larger photometric variations and greater amplitude and frequency variability than shorter period modes in hotter ones \citep{Clemens-1993, Mukadam}. 

Brickhill \citep{Brickhill-1983, Brickhill-1990, Brickhill-1991a, Brickhill-1991b} recognized that the convective flux in DAVs responds on a much shorter timescale than the oscillation periods of their overstable g-modes. He showed that g-modes with $\omega\gtrsim \tau_c^{-1}$ are overstable due to a mechanism he named `convective driving'. Here $\tau_c$ is the effective thermal timescale of the convection zone (CVZ).\footnote{Our paper adopts terminology used by \cite{Goldreich-Wu-I} in their investigation of convective driving.} Brickhill also realized that the amplitude of the flux perturbation that emerges from the photosphere, $\Del F_{ph}$, is diminished by the factor $(1+(\omega\tau_c)^2)^{-1/2}$ relative to the one that enters the bottom of the CVZ, $\Del F_b$. His twin insights explain why overstable modes shift to longer periods as a DAV cools and why longer period modes usually show larger photometric variations. These trends are well-established among DAVs above the red edge.  

By itself,  the overstability criterion, $\omega\tau_c>1$ cannot account for the red edge of the DAV instability strip.  An essential step toward understanding its origin was taken by \cite{Dolez-Vauclair}  in a paper that predated Brickhill's work on convective driving. Their linear nonadiabatic calculations found $\omega>L_{\ell,b}$ to be a necessary condition for the overstability of g-modes in cool DAVs.\footnote{$L_{\ell,b}$ is the Lamb frequency at the bottom of the surface CVZ for modes of angular degree $\ell$.} However, \cite{Dolez-Vauclair} mistakenly attributed overstability to the $\kappa$-mechanism. Later, \cite{Pesnell} noted that the overstability they found was caused by `convective blocking' due to their incorrect assumption of frozen convection.\footnote{That the assumption of frozen convection could lead to mode excitation was well-established prior to its being named convective blocking \citep{Cox1}.}  Convective blocking is the antithesis of convective driving. It is the latter which is responsible for the excitation of overstable g-modes across the entire DAV instabililty strip. The identification by \cite{Dolez-Vauclair} of $\omega>L_{\ell,b}$ as a necessary condition for overstability remains valid under convective driving as presented in the numerical results of \citep{Wu-Goldreich-II}. Thus together, $\omega>1/\tau_c$ and $\omega>L_{\ell,b}$ define the low frequency limit, $\omega_\mathrm{crit}$, for overstable g-modes at each $\Teff$.

Inspired by \cite{Dolez-Vauclair}, Brickhill appears to have accepted $\omega> \Lambb$ as a necessary condition for overstability. However, we are unsure why.  We have been unable to locate a statement in any of Brickhill's papers on DAVs in which he clearly attributes this condition to radiative damping exceeding convective excitation.  Instead, his focus seems to have been on turbulent damping in the convective overshoot region below $\zb$ \citep{Brickhill-1991b}. This is a puzzle, since the calculations by \cite{Dolez-Vauclair} only accounted for radiative damping. 

In our paper, we explain that modes with $\omega<L_{\ell,b}$ suffer enhanced radiative damping that  overwhelms convective driving. We demonstrate that the quasi-adiabatic approximation suffices to account for the enhancement in radiative damping and thus provides a minimally complicated conceptual understanding of the red edge, although it would be ideal to have fully non-adiabatic calculations including turbulent viscosity.  

As a DAV cools, $1/\tau_c$ steeply declines and $L_{\ell,b}$ gradually rises; $\omega_\mathrm{crit}$ reaches a minimum value close to the red edge as defined by photometric surveys of DAVs. Well below the red edge, photospheric flux variations associated with overstable modes are severely diminished; at $\omega_\mathrm{crit}$, $\Del F_{ph}/\Del F_b \sim [(1+(L_{\ell,b}\tau_c)^2]^{-1}\ll 1$.  Moreover, almost nothing is known about DAVs below the red edge because photometry is a poor technique for identifying them.  Although more challenging, spectroscopic observations of bright DAs just below the red edge might succeed in detecting doppler shifts due to the largely horizontal velocities associated with overstable modes. These do not suffer significant reductions between $z_b$ and $z_{ph}$ \citep{van-Kerkwijk}.

The discovery of outbursts in photometric data obtained by the {\it Kepler} satellite from cool DAVs is an exciting recent development \citep{Bell-2015, Hermes, Bell-2016a, Bell-2016b, Hermes-2017}. Individual outbursts recur sporadically with separations of days and emit $10^{33}-10^{34}\erg$ on timescales of hours.  \cite{Hermes} suggested that outbursts may involve nonlinear mode couplings as described by \cite{Wu-Goldreich-IV}. To lowest order, relevant nonlinearities couple an overstable parent mode to a pair of lower frequency stable daughter modes. Photometric amplitudes observed across the DAV instability strip are generally in line with those predicted by calculations of saturation based on 3-mode couplings \citep{Wu-Goldreich-IV}.  

We concur with \cite{Hermes} and associate outbursts with limit cycles exhibited by 3-mode couplings satisfying $\delta\omega<|\gamma_d|$.  Here $\delta\omega\equiv |\omega_p-\omega_{d_1}-\omega_{d_2}|$ is the frequency mismatch between an overstable parent mode (labeled by $p$) and two stable daughter modes (labeled by $d_1$ and $d_2$), and $\gamma_d\equiv (\gamma_{d_1}+\gamma_{d_2})$ is the average damping rate of daughter modes \citep{Wu-Goldreich-IV}. Consistent with observations, this condition is most likely to be satisfied in cool DAVs near the red edge of the instability strip.  The red edge is where the lowest frequency overstable parent modes are found, where the density of these modes per unit frequency is greatest, and where candidate daughter modes have the largest damping rates, $\gamma_d\sim -(n_d\tau_{th,b})^{-1}$.\footnote{Here $n_d$ is the number of radial nodes of the daughter modes and $\tau_{th,b}\sim \tau_c/5$ is the thermal timescale at the bottom of the CVZ, $z_b$.}  Away from the red edge, toward both higher and lower $\Teff$, the lowest frequency for an overstable mode, $\omega_\mathrm{crit}$, increases, and $\delta\omega$ for candidate 3-mode couplings also increases accordingly because density of g-modes per unit frequency grows with higher frequency. Consequently, $\delta\omega<|\gamma_d|$, is most easily achieved near the red edge. 
 
Our paper is arranged as follows. Section~\ref{sec:red-edge} recapitulates the overstability criterion, $\omega>\max(\tau_c^{-1},L_{\ell,b})$, focusing on the physical explanation for the second threshold $\omega>L_{\ell,b}$ which essentially yields the photometric red edge. Section~\ref{sec:burst} argues that parametric instability involving 3-mode couplings that are sufficiently resonant to trigger limit cycles can quantitatively account for the outbursts. We summarize and discuss unresolved problems in Section~\ref{sec:conclusion}. All figures except Figure~\refnew{fig:omegatauc-T} are based on a WD model with $\Teff\approx 10860\,\K$ and $g\approx 10^8\,\mathrm{cm\, s^{-2}}$ that is described in Section~\ref{sec:rededge}.

\section{Red edge}\label{sec:red-edge}

\subsection{Order-of-Magnitude Results}

Gravity waves propagate where $\omega$ is smaller than both the Brunt-Vaisala frequency, $N$, and the Lamb frequency, $\Lamb$ \citep[e.g.][]{Cox}. We denote by $\zw\sim \omega^2 r^2/(g \ell(\ell+1))$ the depth at which $\omega=\Lamb$, and by $z_b$ the depth at the bottom of the surface CVZ. The upper lid of a g-mode's propagation cavity is detached from the CVZ if $\zw>z_b$, or equivalently, $\omega>\Lambb\equiv \Lamb(\zb)$. Otherwise the propagation cavity terminates at $\zb$.

Consider a g-mode with $\zw\gg z_b$.  The mode is evanescent in both the CVZ and the first scale height below $\zb$, and each stores a similar amount of mode energy. Upon compression of the evanescent region, the first scale height, $H_1$, loses heat (entropy) by radiative diffusion, and thus contributes to mode damping, $\gamma_\mathrm{rad}$. Note that $H_1$ dominates $\gamma_\mathrm{rad}$, because the local damping rate is inversely proportional to the local thermal timescale $\tau_\mathrm{th}(z)\sim p H_p/F\appropto z^6$ which grows steeply with depth. Meanwhile entropy lost by the radiative zone gets absorbed into the CVZ and does not escape until after a timescale $\sim \tau_c$. Thus the CVZ contributes to mode excitation.  According to \cite{Goldreich-Wu-I}, the ratio of convective driving to radiative damping is
\begin{equation}
{\gamma_{\mathrm{cvz}}\over\left|\gamma_{\mathrm{rad}}\right|}\sim{2(\omega\tau_c)^2\over 1+(\omega\tau_c)^2}\, ,\,\, (\omega\gg L_{\ell,b})\, .\label{eq:drive-damp-ratio}
\end{equation}

Next we consider a g-mode with $\zw\ll z_b$, or equivalently $\omega\ll L_{\ell,b}$. This mode propagates up to $\zb$, and its effective wavelength, $\lambda_1$, becomes smaller than $H_1$. Radiative diffusion rate per unit volume is proportional to the Laplacian of the temperature perturbation associated with the mode. Therefore $\gamma_\mathrm{rad}$ gets enhanced by a factor $(H_1/\lambda_1)^2 \sim (k_z H_1)^2 \sim (k_h N H_1/\omega)^2\sim (k_h c_s/\omega)^2 \sim (L_{\ell,b}/\omega)^2$.\footnote{We substitute Equation (18) into the second line of Equation (50) in \cite{Goldreich-Wu-I} to get this result. For the purpose of order-of-magnitude estimate, we only retain terms on the order of $k_z^2$ in $\gamma_\mathrm{rad}$ and neglect terms on the order of $k_z /H_1$ and $H_1^{-2}$. We keep all the terms in the quasi-adiabatic calculations described in the subsection \ref{sec:quasi-adiabatic}.} Consequently, radiative damping is enhanced by a factor $(L_{\ell,b}/\omega)^2$ relative to convective driving and 
\begin{equation}
{\gamma_{\mathrm{cvz}}\over\left|\gamma_{\mathrm{rad}}\right|}\sim  {2(\omega\tau_c)^2\over 1+(\omega\tau_c)^2}\left(\omega\over L_{\ell,b}\right)^2\, ,\,\, (\omega\ll L_{\ell,b})\, ,\label{eq:drive-damp-ratio-smallw}
\end{equation}

Hot DAVs have shallow CVZs such that $L_{\ell,b} \ll 1/\tau_c$ and modes with $\omega\ll L_{\ell,b}$ are damped according to Equation~\refnew{eq:drive-damp-ratio-smallw}. Equation~\refnew{eq:drive-damp-ratio} sets the boundary between overstable and damped modes at
\begin{equation}
\omega_\mathrm{crit} \sim {1\over \tau_c}\, , \,\, (L_{\ell,b}\ll 1/\tau_c)\, .\label{eq:omegacrit-hot}
\end{equation}
Cool DAVs have deep CVZs such that $L_{\ell,b}\gg 1/\tau_c$ and modes with $\omega\gg L_{\ell,b}$ are overstable according to Equation~\refnew{eq:drive-damp-ratio}.  Equation~\refnew{eq:drive-damp-ratio-smallw} sets the boundary between overstable and damped modes at 
\begin{equation}
\omega_\mathrm{crit}\sim {L_{\ell,b}\over \sqrt{2}}\, ,\,\, (L_{\ell,b}\gg 1/\tau_c)\, .\label{eq:omegacrit-cool}
\end{equation}

The flux perturbation that emerges from the photosphere is diminished relative to that entering the base of the CVZ following the relation
\begin{equation}\label{eq:dilution}
{\Delta F_{\mathrm{ph}}\over F}\approx {1\over 1-i\omega\tau_c}\left({\Delta F\over F}\right)_{z_b}\, .
\end{equation}
In cool WDs, overstable modes with $\omega\gtrsim \omega_\mathrm{crit}$ all have $\omega\tau_c\gg 1$. Thus they escape detection in photometric surveys. This explains the red edge of the DAV instability strip. It reflects the transition between the two overstability criterions which results from the deepening of the surface CVZ as a WD cools. Existence of overstable modes at $\Teff$ well below the photometric red edge is the key to test our explanation of the red edge. High-quality spectroscopy of a nearby cool WD may reveal velocities associated with an overstable g-mode since velocities do not suffer the reduction effect like the flux perturbation, although such test seems very challenging in practice.\footnote{Based on private communications with Chris Clemens and JJ Hermes.}

\subsection{Comparison with Quasi-Adiabatic Calculations}\label{sec:quasi-adiabatic}
{\it MESA} \citep{Paxton} is employed to produce WD models with total mass, $M\approx 0.60\Msun$, surface gravity, $g\approx 10^8\,\mathrm{cm\, s^{-2}}$, and hydrogen envelope mass, $M_H\sim 10^{-4}M_\mathrm{WD}$. We adopt the ML2 version of the mixing length prescription, and adjust the mixing length, $\alpha$, such that entropy levels at the bottom of the CVZ agree with those in Figure~(1) of \cite{Tremblay-MLT}.  Our WD models are input to {\it gyre} \citep{Townsend} to calculate adiabatic eigen-frequencies and eigen-functions for g-modes. The quasi-adiabatic technique as described in \cite{Goldreich-Wu-I} is used to calculate convective driving and radiative damping rates. We adopt the full expression for the quasi-adiabatic flux perturbation in Equation (10) of \cite{Goldreich-Wu-I}, and then we apply their equation (50) to calculate the driving and damping rates. Here we focus on comparing the order-of-magnitude results with those obtained from quasi-adiabatic calculations.

\begin{figure}
\includegraphics[width=0.99\linewidth]{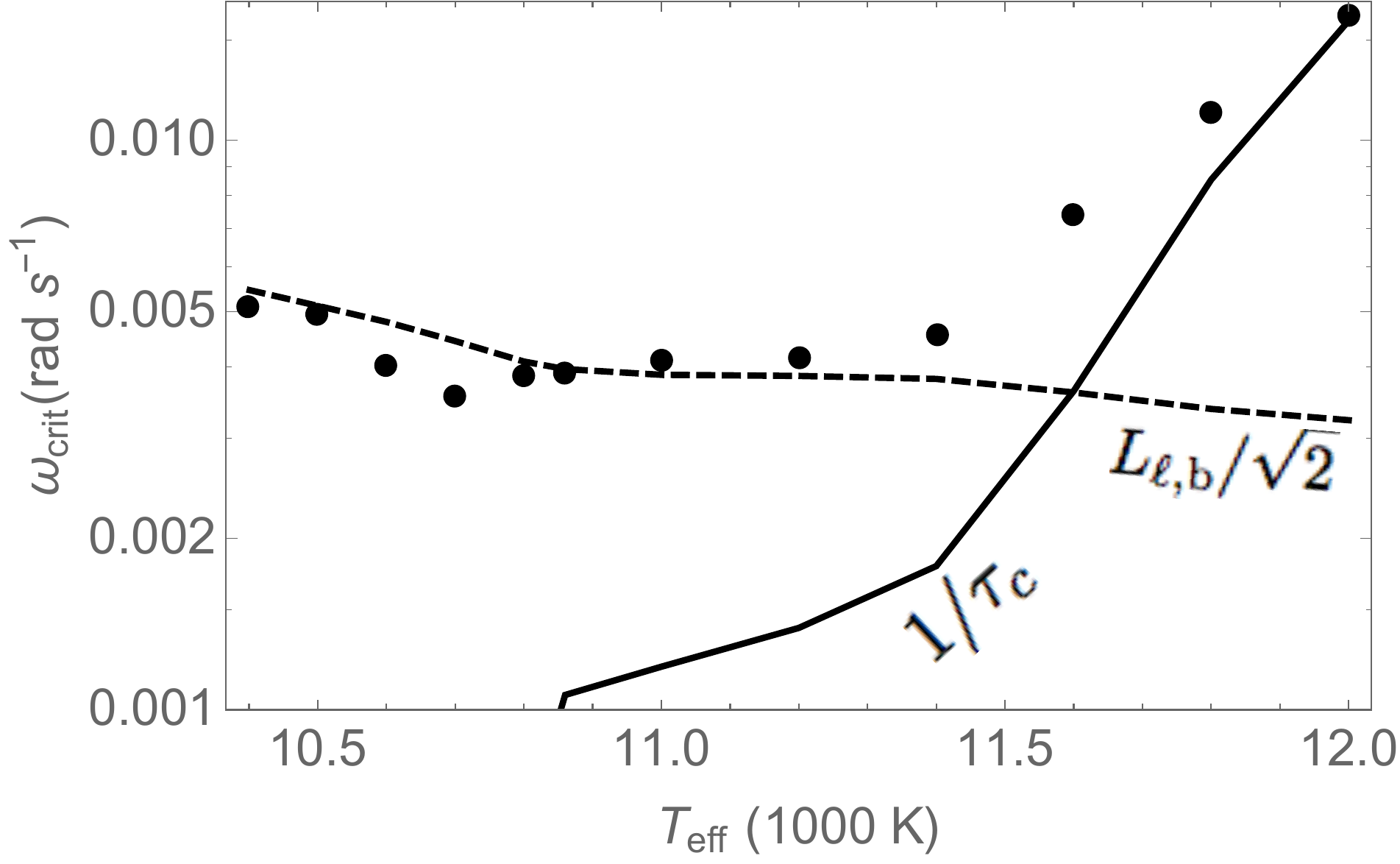}
\caption{\label{fig:omegacrit-Teff}  The black dots show the radian frequency for marginally overstable modes, $\omega_\mathrm{crit}$, versus $\Teff$, based on quasi-adiabatic calculations. The solid curve shows $1/\tau_c$, the overstability criterion suitable for hot DAVs. The dashed one shows $L_{\ell,b}/\sqrt{2}$, the overstability criterion suitable for cool WDs.    }
\end{figure}

Figure~\refnew{fig:omegacrit-Teff} shows that angular frequencies of marginally overstable g-modes with $\ell=1$ as a function of $\Teff$ exhibit a gradual transition between the two limiting cases, Equations~\refnew{eq:omegacrit-hot} and \refnew{eq:omegacrit-cool}.

\begin{figure}
\includegraphics[width=0.99\linewidth]{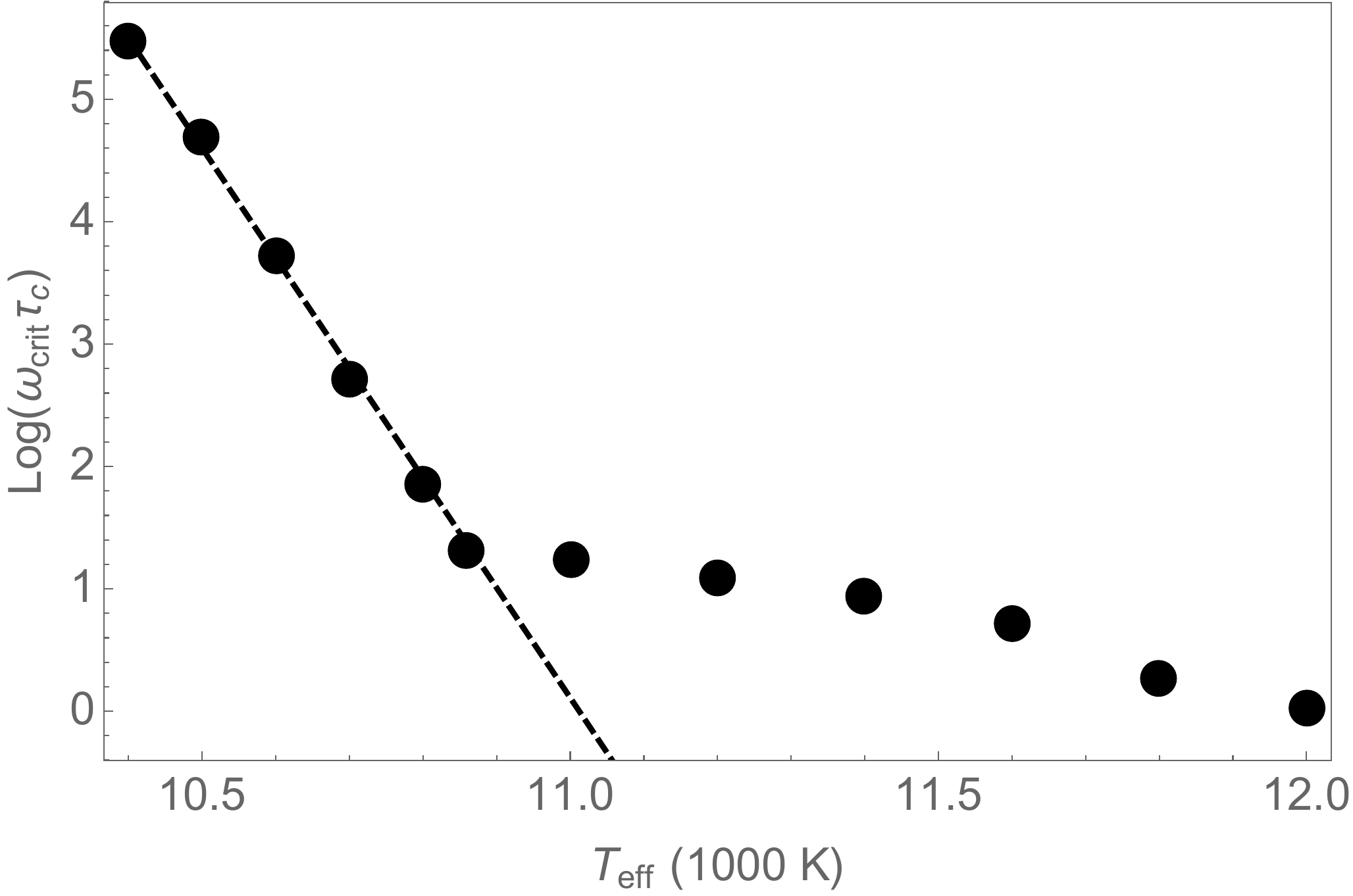}
\caption{\label{fig:omegatauc-T}  The reduction factor for the photometric flux perturbation for marginally overstable modes, $\omega_\mathrm{crit}\tau_c$, versus $\Teff$. The black dots are based on our quasi-adiabatic calculations using WD models made by MESA as detailed in Section~\ref{sec:rededge}. The dashed line plots the fitting formula, Equation~\refnew{eq:omegatauc-Teff-fit}. The observational photometric red edge is $\sim 10800\,\mathrm{K}$ where $\omega_\mathrm{crit}\tau_c\sim 10$. The reduction factor increases exponentially with lower temperature, making overstable modes invisible below the red edge. }
\end{figure}

Figure~\refnew{fig:omegatauc-T} plots the reduction factor for the photometric variation, $\omega_\mathrm{crit}\tau_c$. It is on the order of a few above the observational red edge, and increases exponentially with lower $\Teff$ below the red edge. The dashed line show our fitting formula,
\begin{equation}\label{eq:omegatauc-Teff-fit}
\omega_\mathrm{crit}\tau_c\sim \exp\left[(11012\,\mathrm{K} - \Teff)\over 111\,\mathrm{K}\right]\, ,\, (\ell=1)\, .
\end{equation}
Therefore, overstable modes become invisible below the red edge.


$\Teff$ at the red edge depends sensitively upon $\zb$, the depth the surface CVZ. But $\zb$ has a sensitive dependence on the mixing length parameter, $\alpha$.  This warrents a short detour to explain how $\alpha$ affects $\zb$. Except for the top scale height right below the photosphere, the bulk of the CVZ is nearly isentropic, i.e., $ds/dz\sim 0$, so the specific entropy is well-approximated by $s_b$. Given $\Teff$ and $g$ at the photosphere, a smaller $\alpha$ corresponds to a smaller convection efficiency and thus yields a smaller $ds/dz$ in the top scale height, and therefore a lower $s_b$. At fixed density, smaller entropy corresponds to lower temperature and higher opacity, and thus a deeper surface CVZ. WD models at the same $\Teff$ but with different choices of $\alpha$ could have completely different depths of CVZs and thus would predict very different $\omega_\mathrm{crit}$. The $\alpha$ in our WD models at different $\Teff$ are chosen so that the $s_b$ agrees with those in \cite{Tremblay-MLT} who calibrated $s_b$ by observational spectra of WDs.  

\subsection{At the Red Edge}\label{sec:rededge}
Material in this subsection is based on a particular DA WD model at $\Teff\approx 10860\,\mathrm{K}$ and $g\approx 10^8\,\mathrm{cm\, s^{-2}}$ with the longest period overstable dipole modes at $P\approx 20\min$. At the base of the CVZ, the specific entropy and pressure are $s_b\sim 2.6\times 10^9\,\mathrm{erg\, g^{-1}\, K^{-1}}$ and $p_b\approx 7.5\times 10^8\,\mathrm{dyn\, cm^{-2}}$, the conventional thermal timescale, $\tau_\mathrm{th,b}\equiv \int^{z_b} dz {c_p \rho T/F}\approx 5.13\,\mathrm{min}$ as compared to the effective thermal timescale, $\tau_c\approx 15.6\,\mathrm{min}$. Radiative damping is concentrated within the first scale height below the CVZ, $z_b<z<z_b+H_1$. At the red edge, the lowest frequency of overstable modes satisfies $\omega_\mathrm{crit}\tau_\mathrm{th}\approx 1.5$ at $z_b$, and $\omega_\mathrm{crit}\tau_\mathrm{th}\sim 7.2$ at $z_b+H_1$. Therefore, the quasi-adiabatic condition, $\omega\tau_\mathrm{th}\gg 1$, is marginal for overstable modes at the red edge. Because $\omega_\mathrm{crit}\tau_c$ grows exponentially with lower $\Teff$ below the red edge and $\tau_c\sim 5\tau_\mathrm{th}(z_b)$, the quasi-adiabatic condition is well-satisfied for overstable modes in DAVs that are cooler than those at the red edge. Ideally, non-adiabatic calculations that include turbulent viscosity should be carried out to determine the theoretical red edge temperature, $T_\mathrm{red}$. However, as described in the previous subsection, the choice of the mixing length $\alpha$ already inserts considerable uncertainty in $T_\mathrm{red}$. Quasi-adiabatic approximations have the virtue of technical simplicity while being adequate to explain the existence of the red edge.

\begin{figure}
\includegraphics[width=0.9\linewidth]{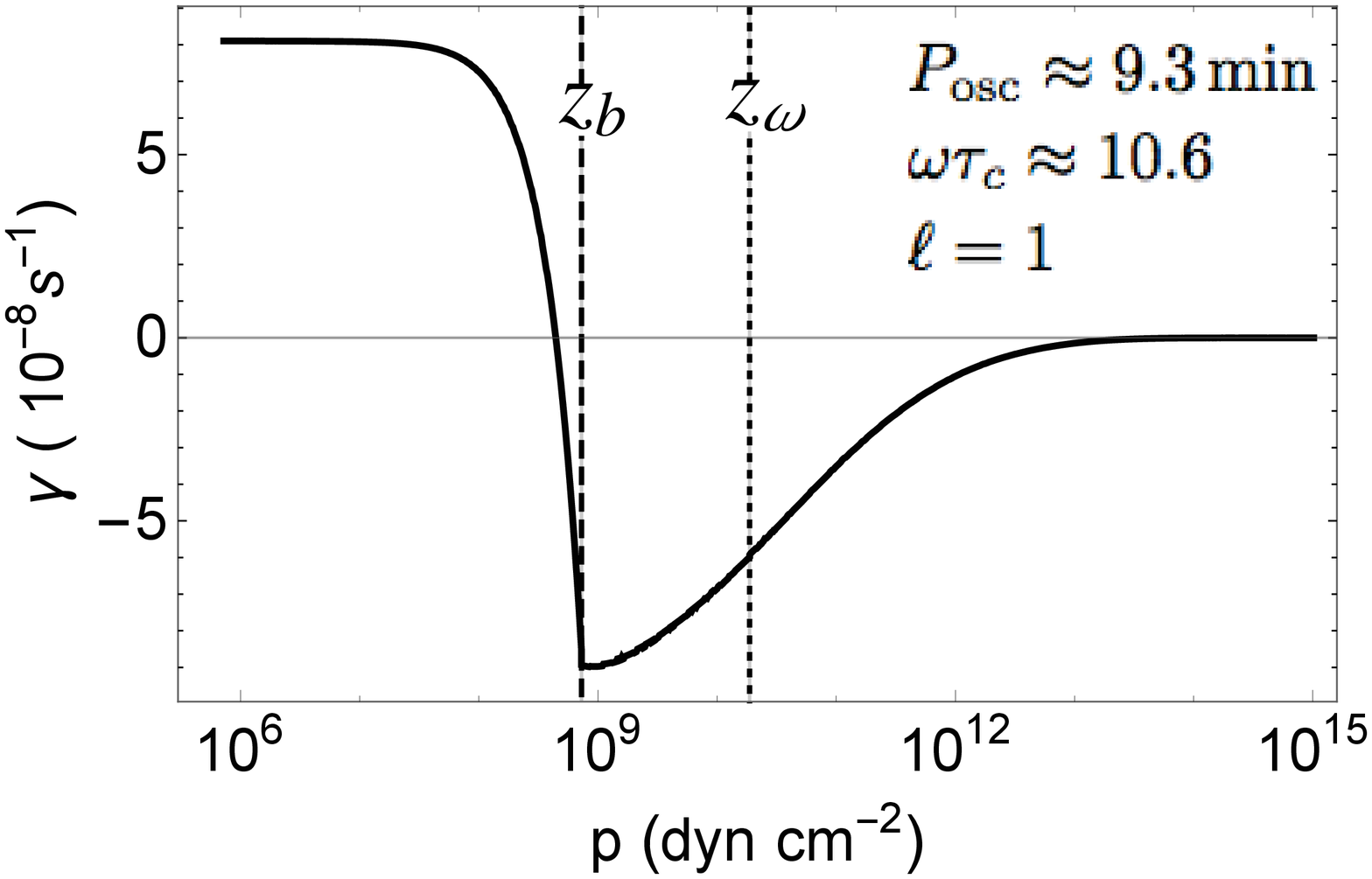}
\includegraphics[width=0.9\linewidth]{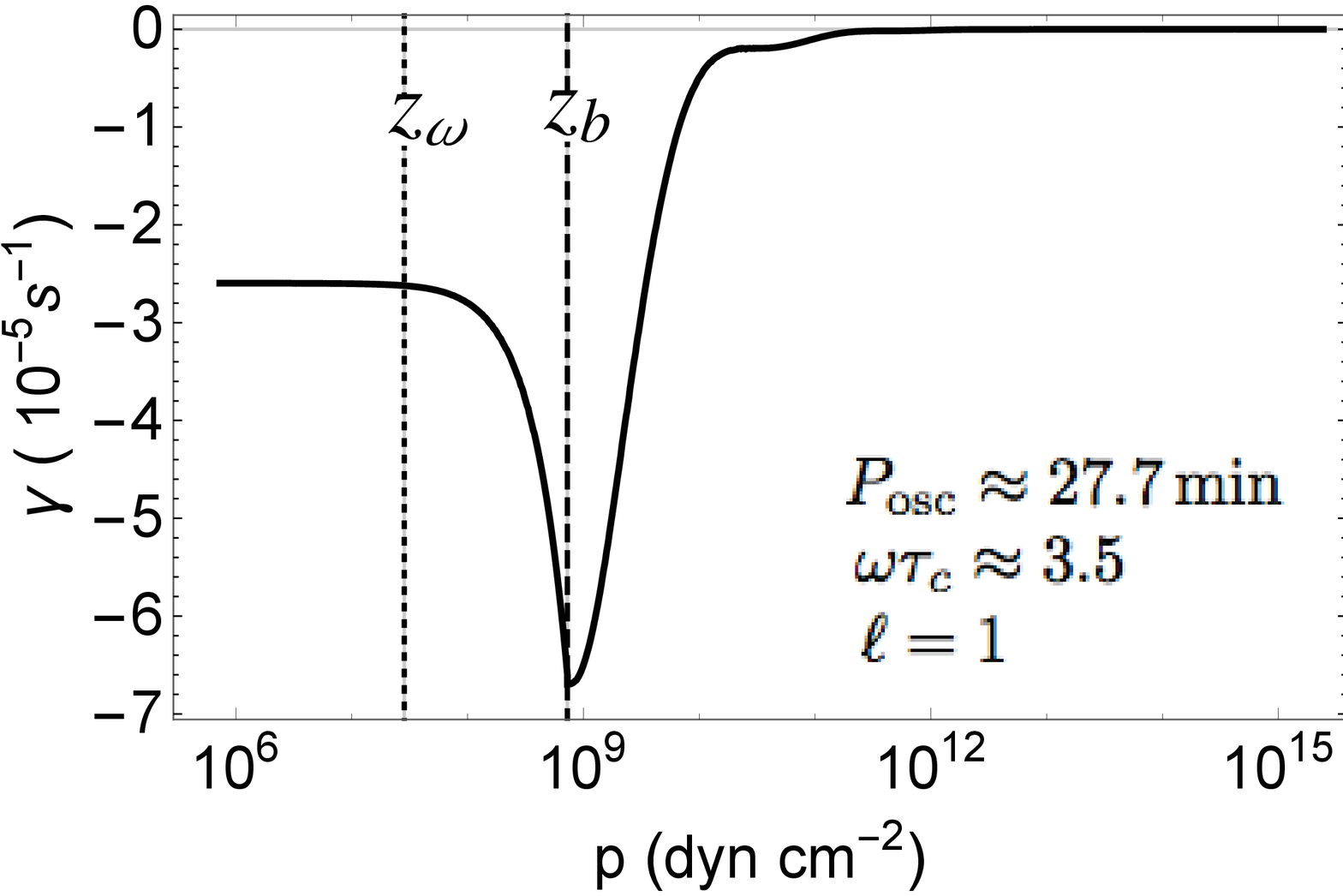}
\caption{\label{fig:WI-compare} Work integrals for two $\ell=1$ g-modes in our MESA WD model with $\Teff\approx 10860\K$. The upper panel and lower panels demonstrate that the mode with $\zw\gg\zb$ is overstable, whereas the mode with $\zw\ll \zb$ is damped. Note that $\gamma$ at pressure $p$ is the net contribution to damping and driving from the region between the stellar center up to pressure $p$. The total damping or driving rate is given by the left end of the black solid curve in each panel. Note that quasi-adiabatic approximation for the mode shown in the lower panel is marginal.}
\end{figure}

\begin{figure}
\includegraphics[width=0.99\linewidth]{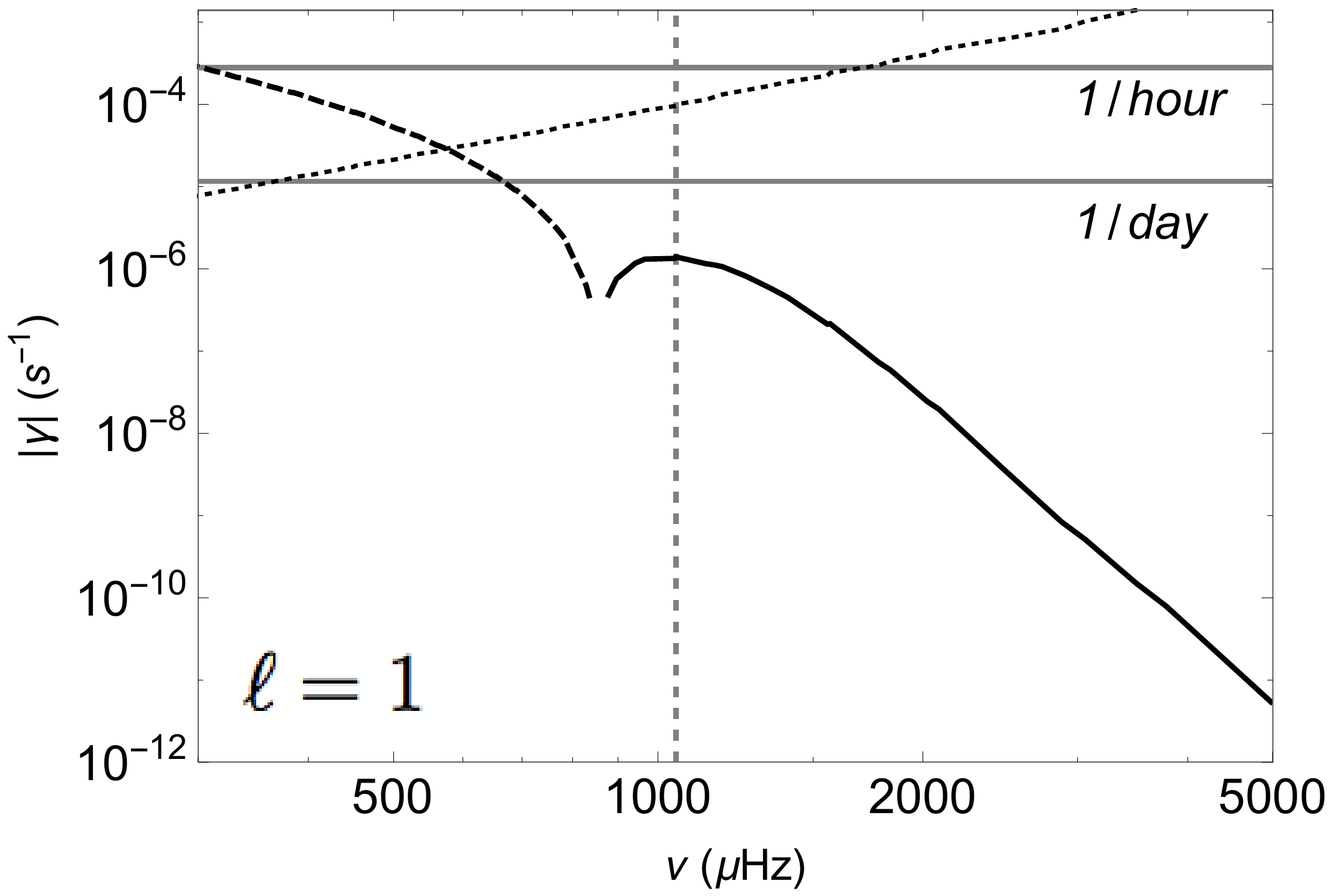} 
\includegraphics[width=0.99\linewidth]{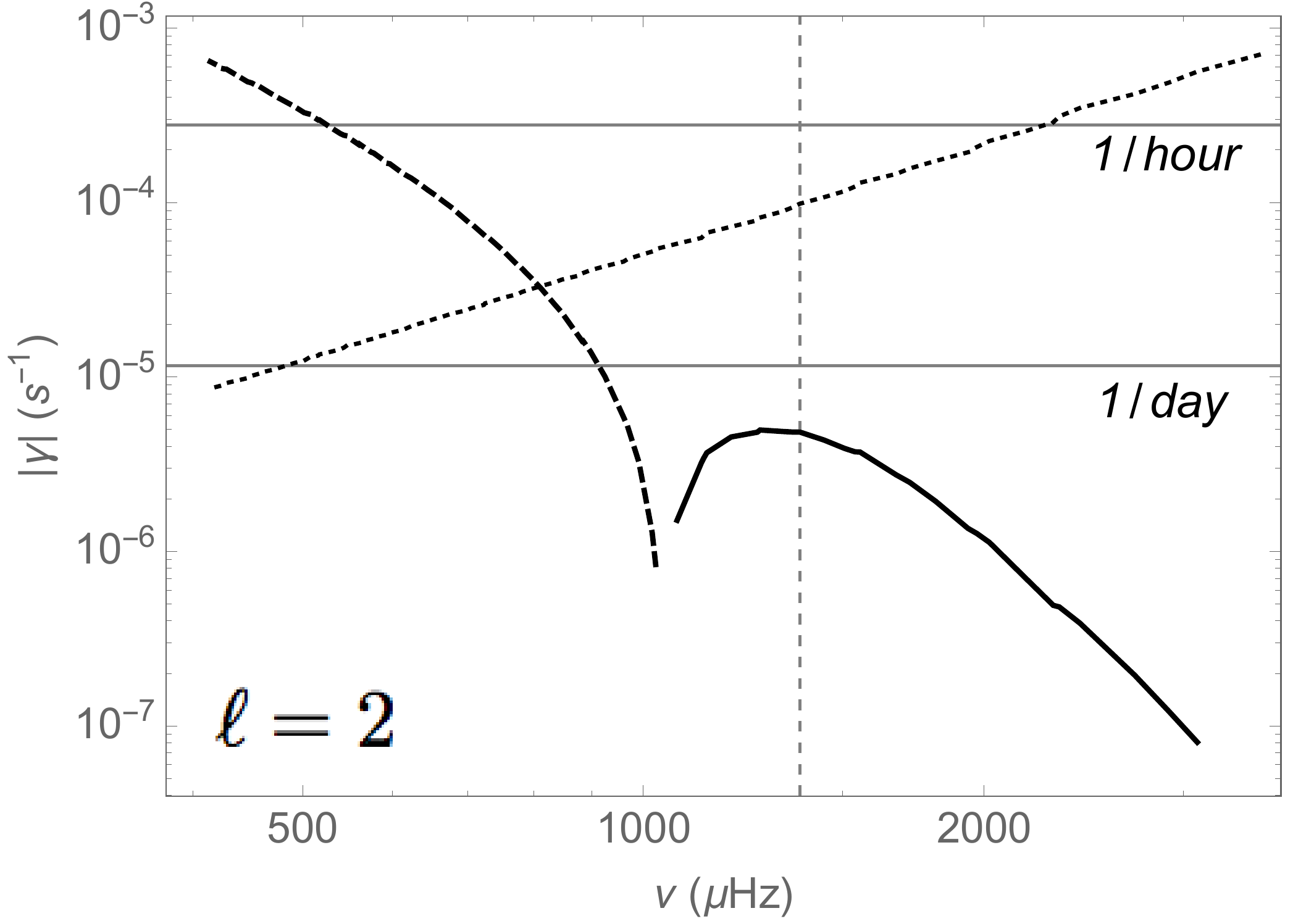} 
\includegraphics[width=0.99\linewidth]{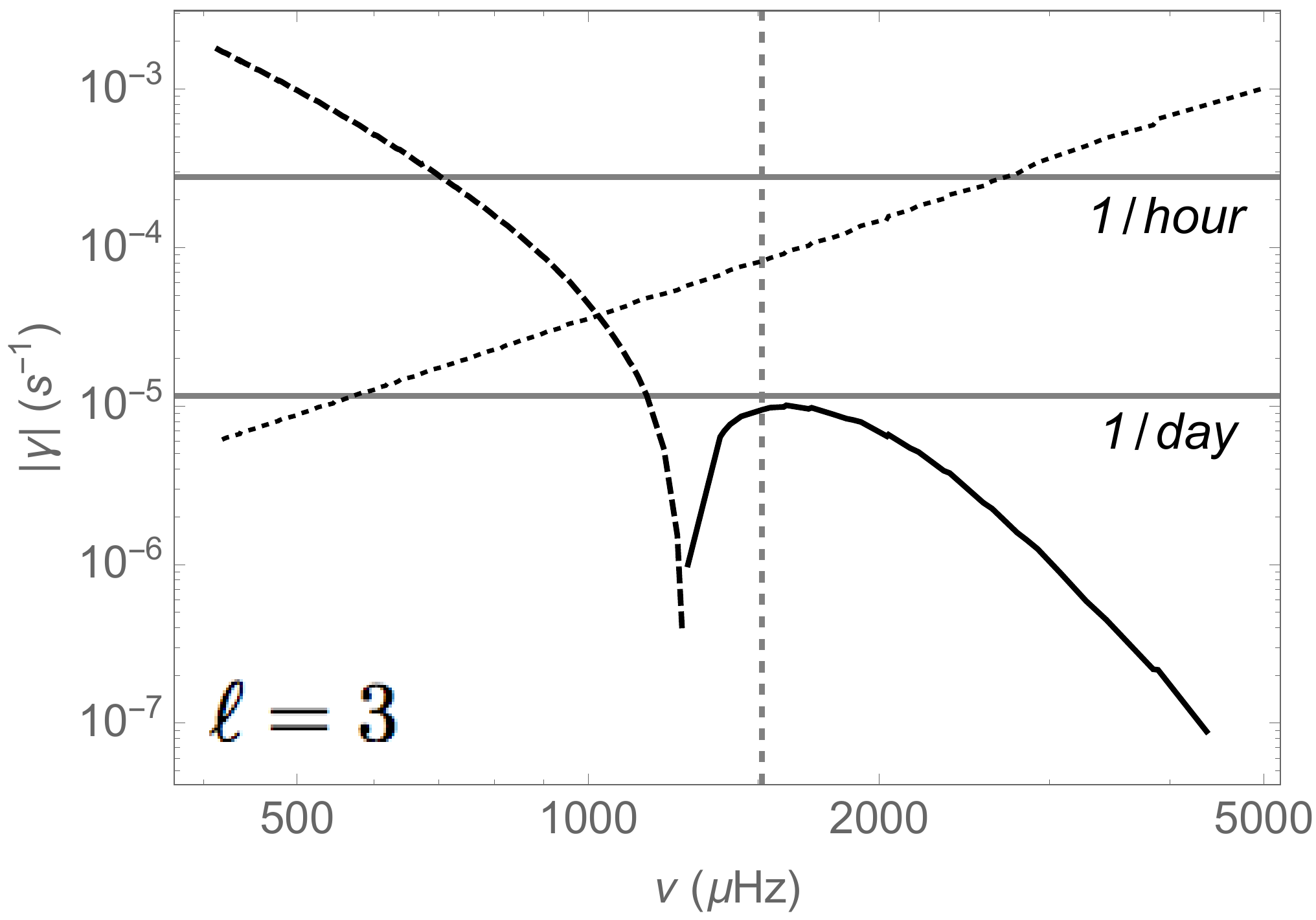} 
\caption{\label{fig:timescale} Linear growth and damping rates for g-modes with $\ell=1,\,\, 2,\,\, 3$. Solid curves correspond to overstable modes with $\gamma>0$, whereas dashed curves apply to damped modes with $\gamma<0$. Dotted lines indicate $\omega/(\pi n)$, the reciprocal of the time a g-wave takes to travel across its cavity.  A mode transitions to a traveling wave if its damping rate is greater than $\omega/(\pi n)$. 
In the same WD model at $\Teff\approx 10860\,\mathrm{K}$, the lowest frequency of an overstable mode rises with increasing $\ell$. Corresponding values of $\omega_\mathrm{crit}\tau_c$ are approximately, $5$ for $\ell =1$, $6$ for $\ell =2$, and $7$ for $\ell =3$. Peak driving rates and associated frequencies, $(\gamma_\mathrm{pk},\nu_\mathrm{pk}, \omega_\mathrm{pk}\tau_c)$, are $(1.4\times 10^{-6}\,\,\mathrm{s^{-1}}, 1048\,\,\mu\mathrm{Hz}, 6.2)$ at $\ell=1$, $(5.1\times 10^{-6}\,\,\mathrm{s^{-1}}, 1376\,\,\mu\mathrm{Hz}, 8.1)$ at $\ell=2$, and $(1.1\times 10^{-5}\,\mathrm{s^{-1}}, 1604\,\mu\mathrm{Hz}, 9.4)$ at $\ell=3$. Horizontal grid lines label $1/(\mathrm{day})$ and $1/(\mathrm{hour})$. Dashed vertical lines mark the overstable mode with maximal growth rate, $\gamma_\mathrm{pk}$. }
\end{figure}

Work integrals for an overstable dipole mode with $\zw\gg\zb$ and a damped diple mode with $\zw\ll \zb$ are displayed in Figure~\refnew{fig:WI-compare}.  
Results from quasi-adiabatic calculations of linear growth and damping rates for g-modes with $\ell=1,\,\, 2,\,\, 3$ are summarized in Figure~\refnew{fig:timescale}. We will take use of these results for calculating the 3-mode couplings in the next part of the paper.

\subsection{Comparison with Previous Work}\label{sec:sub-comparison}
As described in Section \ref{sec:intro}, nonadiabatic calculations by \cite{Dolez-Vauclair} first identified $\omega> \Lambb$ as a necessary condition for the overstability of g-modes in DAVs. Not long after, \cite{Brickhill-1983}, realized that \cite{Dolez-Vauclair} erred when they attributed mode excitation to the $\kappa$-mechanism. Because the response time of convection in DAVs is shorter than the periods of overstable modes, he proposed a new excitation mechanism that he named `convective driving'.  Brickhill seems to have accepted $\omega> \Lambb$ as a necessary condition for mode overstability although it is not obvious to us why he did so. In subsequent papers, Brickhill investigated damping by turbulent viscosity in the region of convective overshoot below $z_b$ but to our knowledge, he never clearly implicated radiative damping as the culprit.  

A quasi-adiabatic stability analysis by \cite{Goldreich-Wu-I} confirmed Brickhill's results regarding convective excitation and radiative damping for modes with $\omega\gg\Lambb$.  More significantly, non-adiabatic 
stability calculations by \cite{Wu-Goldreich-II} that include convective driving obtained a similar behavior for $\omega_\mathrm{crit}$ as a function of $\Teff$ to that observed by \cite{Dolez-Vauclair}. The results of the current investigation resemble those obtained by \cite{Wu-Goldreich-II} in the sense that as a DAV cools $\omega_\mathrm{crit}$ shows a steep decline above the red edge and a gradual rise below it. The principal difference between \cite{Wu-Goldreich-II} and our current ones is that the former find the theoretical red edge to occur at a $\Teff$ approximately $1000\K$ greater than we do. It is likely that this discrepancy stems from differences in the parameterization of convection.

\section{Outbursts}\label{sec:burst}
Photometric outbursts from 6 cool DAVs were discovered in data collected by the Kepler satellite \citep{Bell-2015,Hermes, Bell-2016a, Bell-2016b}.\footnote{According to Hermes (private communication), the number of DAVs exhibiting outbursts currently stands at 12.}  Individual outbursts release $\sim 10^{33}-10^{34}\erg$ over several hours and recur aperiodically with separations of days. \cite{Hermes} suggested that outbursts may result from parametric instability \citep{Dziembowski,Wu-Goldreich-IV}.  

The simplest case of parametric instability involves the nonlinear coupling of an overstable parent mode to two damped daughter modes. It requires the parent mode's amplitude, $A_p$, to exceed the threshold value 
\begin{equation}
|A_p|^2=\frac{\gamma_{d_1}\gamma_{d_2}}{18\kappa^2\omega_{d_1}\omega_{d_2}}\left[1+\left(\frac{2\delta\omega}{\omega_{d_1}+\omega_{d_2}}\right)^2\right]\, ,
\label{eq:ampth}
\end{equation}
where $\delta\omega\equiv |\omega_p-\omega_{d_1}-\omega_{d_2}|$ is the frequency mismatch and $\kappa$ is the nonlinear coupling coefficient as defined in Equation~(9) of \cite{Wu-Goldreich-IV}. Here subscripts $p$, $d_1$, $d_2$ denote quantities pertaining to the parent and daughter modes. For $\delta\omega> |\gamma_d|\equiv |(\gamma_{d_1}+\gamma_{d_2})/2|$, parametric instability settles down on an equilibrium state in which the parent and daughter modes maintain constant amplitudes and the star's luminosity is slightly elevated. More stringent resonant couplings with $\delta\omega<|\gamma_d|$ yield limit cycles. We propose that they generate photometric outbursts. 

Individual cycles consist of three stages. In the first, the parent mode's energy exponentiates. The second stage begins once the parent mode's amplitude has grown sufficiently such that the rate at which energy is transferred to its daughter modes overcomes their linear damping rates and their amplitudes begin to grow. At this point, the parent mode's energy is 
\begin{equation}\label{eq:Eth}
E_{\mathrm{th}}\sim \frac{\gamma_{d_1}\gamma_{d_2}}{18 \kappa^2\omega_{d_1}\omega_{d_2}}\, .
\end{equation}
Subsequently, the parent mode's energy continues growing until eventually it drains into the daughter modes 
as fast as convective driving can replenish it.\footnote{At maximum, the parent mode's energy may significantly exceed $E_\mathrm{th}$ \citep{Kumar-Goodman}.} This initiates the third and final stage in the limit cycle during which the parent mode's energy undergoes a precipitous decline. Once it drops below the threshold, the parent mode becomes too weak to feed the daughter modes, and thus the daughter modes'  mechanical energies also decay, since radiative damping converts them into thermal energy. The net result is the conversion of the peak mechanical energy of the parent mode to a pulse of thermal energy.  An outburst arises as the pulse of thermal energy leaks through the photosphere. This completes one cycle.

Growth of the parent mode's energy in the first and second stages occurs on timescale $\sim\gamma_p^{-1}$, and may not be visible; the parent mode's propagation cavity could be deeply buried below the CVZ such that $\omega_p\tau_c\gg 1$ and its photometric flux perturbation is severely reduced  (see equation~\ref{eq:dilution}), or the parent mode might have $\ell>1$ such that the angularly integrated flux perturbation is reduced. These stages corresponds to the quiescent interval between consecutive outbursts. The third stage, conversion of the parent mode's peak energy into heat by the daughter modes lasts for $\sim \gamma_d^{-1}$. The duration of an outburst is set by the maximum of $\gamma_d^{-1}$, $\tau_\mathrm{th}$ at the depth where the thermal pulse is deposited, and $\tau_c$.  The peak energy of the parent mode sets an upper limit to the energy emitted in an outburst.  

As shown by Equation~(12) in \cite{Wu-Goldreich-IV},  magnitudes of 3-mode coupling coefficients are bounded from above by 
\begin{equation}\label{eq:kappa-max}
|\kappa|_{\mathrm{max}}\sim (n_p^3 L_* \tau_{\omega_p})^{-1/2}\, .
\end{equation}
Here $n_p$ is the number of radial nodes in the parent mode, $L_*$ the stellar luminosity, and $\tau_{\omega_p}$ the thermal timescale at depth $z_{\omega_p}$. Thus the lower bound for the threshold energy is
\begin{equation}\label{eq:Ep-min}
E_\mathrm{th,min}\sim \frac{\gamma_{d_1}\gamma_{d_2}}{18 \omega_{d_1}\omega_{d_2}}\, n_p^3\, L_* \,\tau_{\omega_p}\, .
\end{equation}

\begin{figure}
\includegraphics[width=0.99\linewidth]{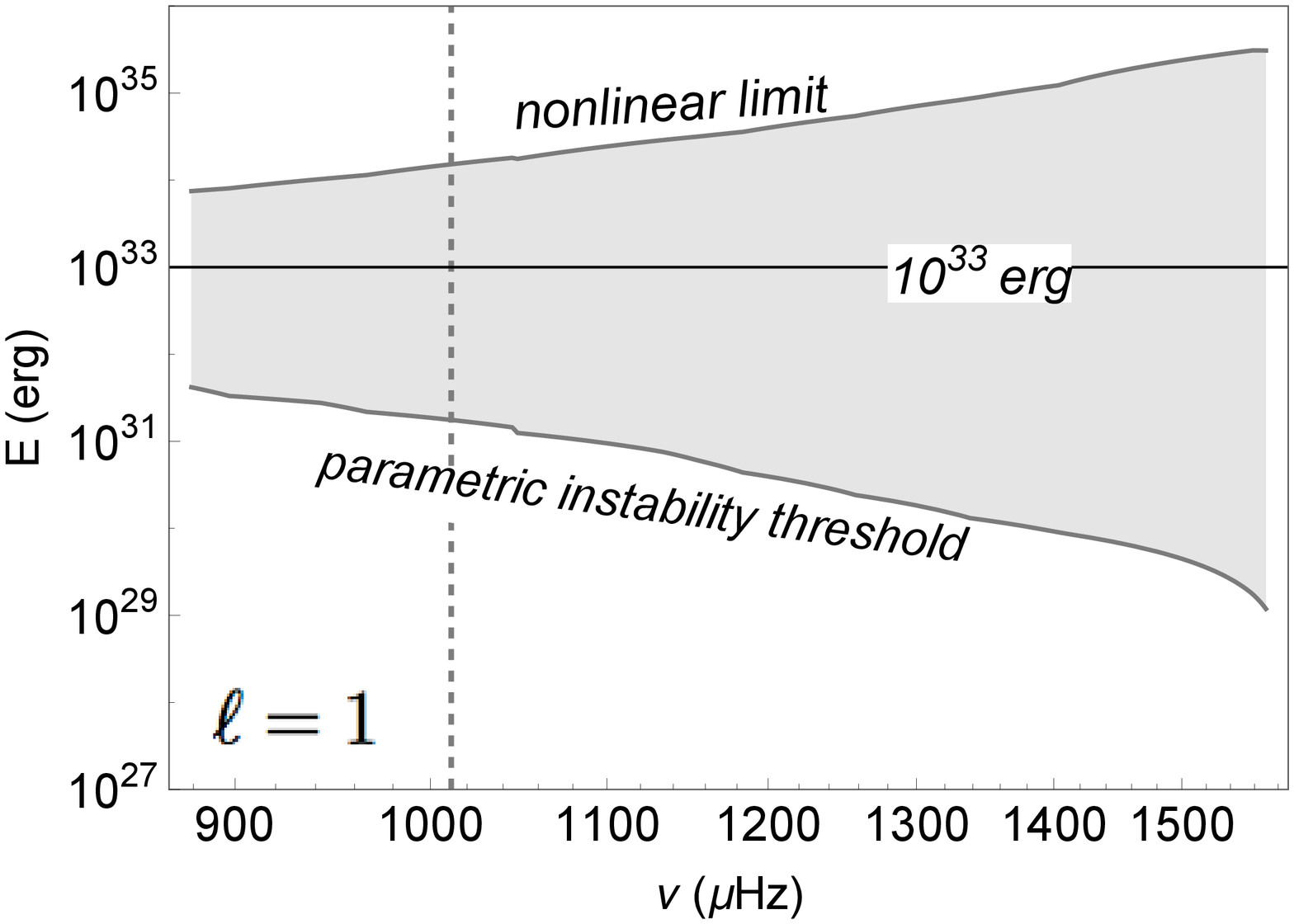}
\includegraphics[width=0.99\linewidth]{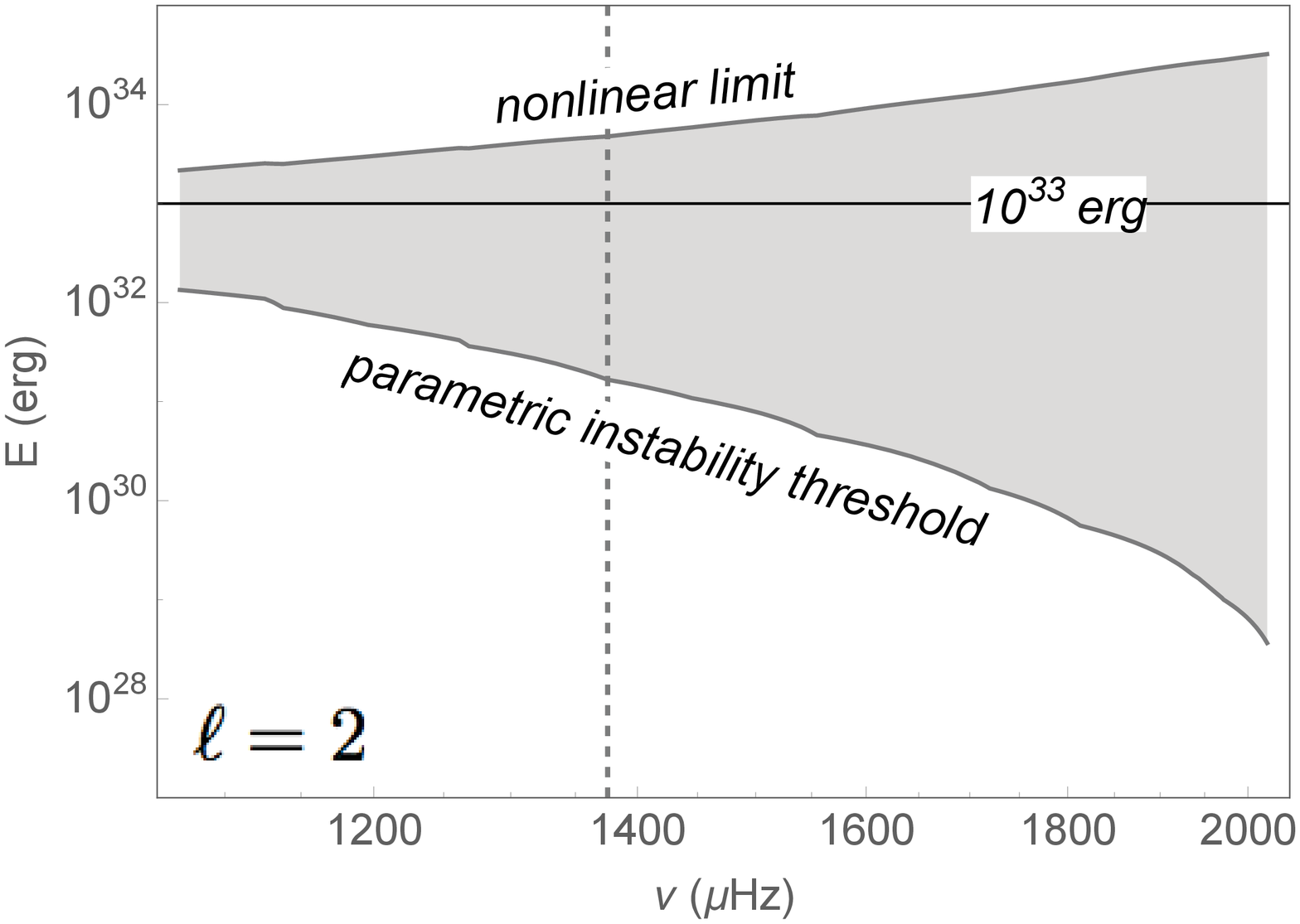}
\includegraphics[width=0.99\linewidth]{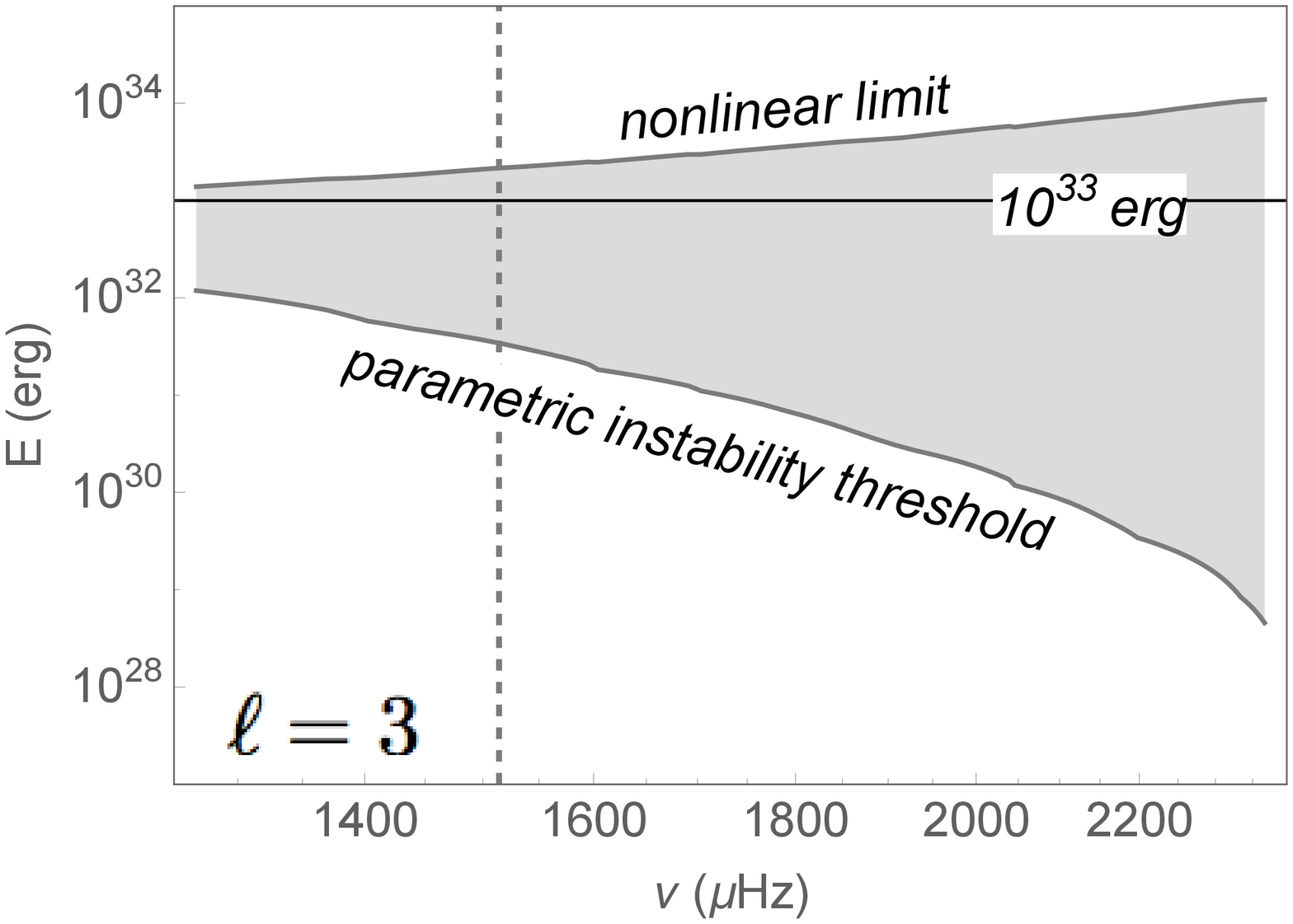}
\caption{\label{fig:E-range} Energy range for prospective parent modes with $\ell=1,\,\, 2,\,\, 3$. The vertical grid lines label the frequencies of the prospective overstable parent modes. These are the modes with maximal growth rates at each $\ell$ (cf. Figure~\refnew{fig:timescale}). Horizontal lines label the typical energy released per outburst.}
\end{figure}

Nonlinear effects constrain $\delta p/p<1$ everywhere throughout the star.  Since $\delta p/p$ increases outward as the background density drops, the upper bound on the parent mode's energy corresponds to $(\delta p/p)_{\mathrm{ph}}=1$.\footnote{In reality, it is difficult to determine the energy of the parent mode, because more descendant modes probably also participate in the energy cascade \citep{Kumar-Goodman}. \cite{Arras} propose that energy cascades from parent mode to descendant modes in a way similar to the Kolmogorov turbulent energy cascade. However, \cite{Brink} conclude that a small fraction of 3-mode couplings dominate the energy cascade and that the energy spectrum differs from Kolmogorov.} Figure~\refnew{fig:E-range} bounds the energy of prospective parent modes between an upper limit corresponding to $(\delta p/p)_{\mathrm{ph}}=1$ and the lower limit $E_\mathrm{th,min}$. Note that $(\delta p/p)^2_{\mathrm{ph}}=1$ corresponds to $E_p\sim n_p\tau_{\omega_p}L_*$  \citep{Goldreich-Wu-I} so $E_p$ rises with increasing $\omega_p$ because $\tau_{\omega_p}$ grows much faster than $n_p$ decreases.

Next, we examine timescales for limit cycles to assess how they compare to recurrence rates and durations of outbursts. Figure~\refnew{fig:timescale} plots driving and damping rates, $\gamma$, for overstable and damped modes with $\ell =1,\, 2,\, 3$.  For each $\ell$, we define $\omega_\mathrm{pk}$ as the frequency at which the peak driving rate, $\gamma_\mathrm{pk}$, occurs.  Note that $\omega_\mathrm{pk}$ is only slightly larger than the minimum frequency for an overstable mode, and also that  $\omega_\mathrm{pk}\tau_c$ increases slowly with $\ell$, running from $6$ at $\ell=1$ through $8$ at $\ell=2$ to $9$ at $\ell=3$. More importantly, values of peak growth rates, $\gamma_\mathrm{pk}$ rise steeply with increasing $\ell$ as do the damping rates of prospective daughter modes.   

We select a parent mode with $\omega\approx\omega_\mathrm{pk}$ along with a matched pair of near resonant daughter modes with $\omega\approx \omega_\mathrm{pk}/2$. To quantitatively account for observed outbursts, these must satisfy
\begin{eqnarray}
\gamma_\mathrm{pk} &\sim&(\mathrm{day})^{-1}\, ,  \label{eq:gamma-p-d}\\
\gamma_d&\sim &(\mathrm{hour})^{-1}\, , \nonumber\\
\delta\omega&<&\gamma_d\, .\nonumber
\end{eqnarray}
Horizontal lines in Figure~\refnew{fig:timescale} label levels at $1/\mathrm{day}$ and $1/\mathrm{hour}$. 

Modes with $\omega\approx \omega_\mathrm{pk}$ are particularly well-suited for driving limit cycles. Radial orders of high $n$ g-modes roughly obey \footnote{See Figure 4 in \cite{Goldreich-Wu-I}.}
\begin{equation}
n \propto {[\ell(\ell+1)]^{1/2}\over \omega}\, .
\end{equation}
Thus the density of modes per unit frequency, $dn/d\omega\propto [\ell(\ell+1)]^{1/2}/\omega^2$, is larger at higher $\ell$ and lower $\omega$.  This is beneficial for finding combinations of near resonant parent and daughter modes with $\delta\omega<|\gamma_d|$ that are suitable for giving rise to limit cycles and thus producing outbursts.

Choosing potential parent modes likely to be responsible for the outbursts involves an interesting set of tradeoffs. The price for short limit cycles and short pulses is a smaller upper bound on peak parent mode energy. Currently, outbursts are only observed through flux increases and flux increases due to frequent short pulses may rival those from rarer long pulses although the latter have larger fluences.

\begin{figure}
\includegraphics[width=0.99\linewidth]{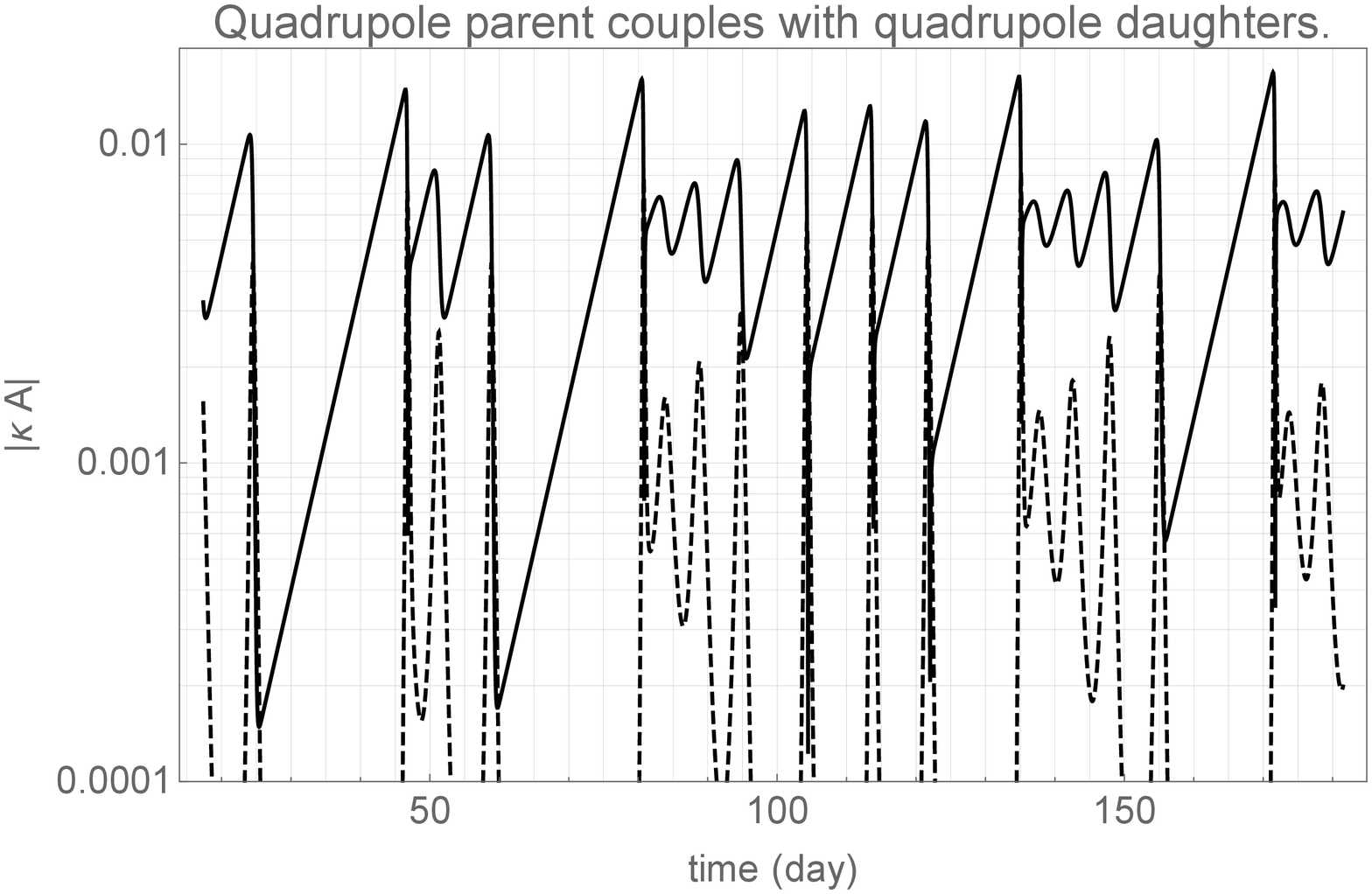}
\includegraphics[width=0.99\linewidth]{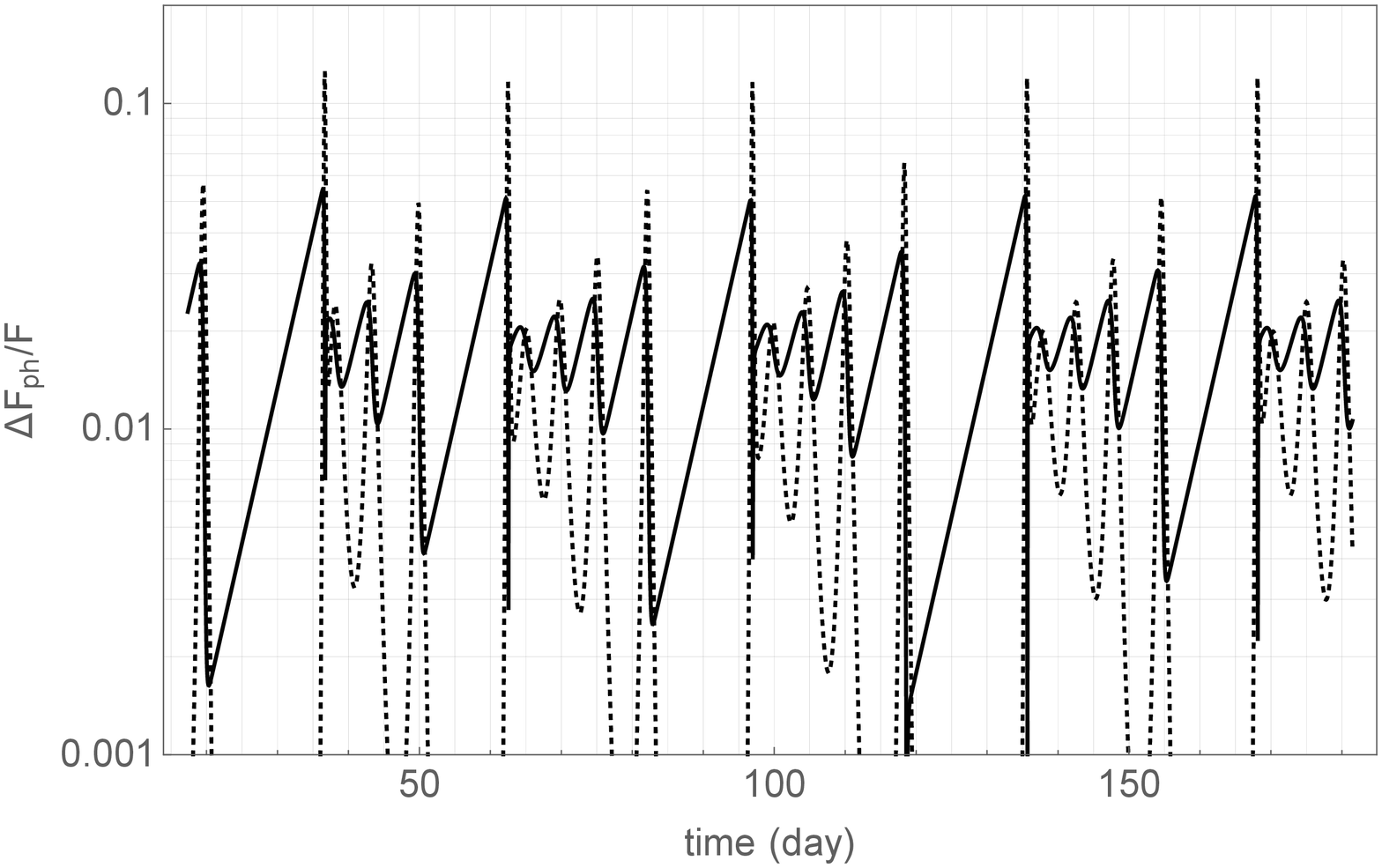}
\caption{\label{fig:outburst-22} Limit cycles involving a quadrupole parent mode and two quadrupole daughter modes with oscillation periods $P_p\approx 12.1145\,\mathrm{min}$, $P_{d_1}\approx 23.8249\,\mathrm{min}$, and $P_{d_2}\approx 24.6537\,\mathrm{min}$. Radial orders are $n_p=28$, $n_{d_1}=58$ and $n_{d_2}=59$. Rates of driving and damping are $\gamma_p\approx 5.1017\times 10^{-6}\,\mathrm{s^{-1}}$, $\gamma_{d_1}\approx -9.6923\times 10^{-5}\,\mathrm{s^{-1}}$, and $\gamma_{d_2}\approx - 1.1678\times 10^{-4}\,\mathrm{s^{-1}}$. The frequency mismatch, $\delta\omega\equiv \omega_p-\omega_{d_1}-\omega_{d_2} \approx 1.17736\times 10^{-6} \,\mathrm{rad\, s^{-1}}$. The modal frequencies, radial orders, and driving/damping rates are calculated based on the WD model at $\Teff\approx 10860\K$ and $g\approx 10^8\,\mathrm{cm\, s^{-1}}$. The upper panel plots the dimensionless $|\kappa A|$ for the parent mode (solid line) and the first daughter mode (dashed line). The lower panel plots their photometric flux perturbations, assuming the peak $|\kappa A_p|\approx 0.015$ in the upper panel corresponds to $\approx 10^{33}\,\mathrm{erg}$ for the parent mode.  }
\end{figure}

\begin{figure}
\includegraphics[width=0.99\linewidth]{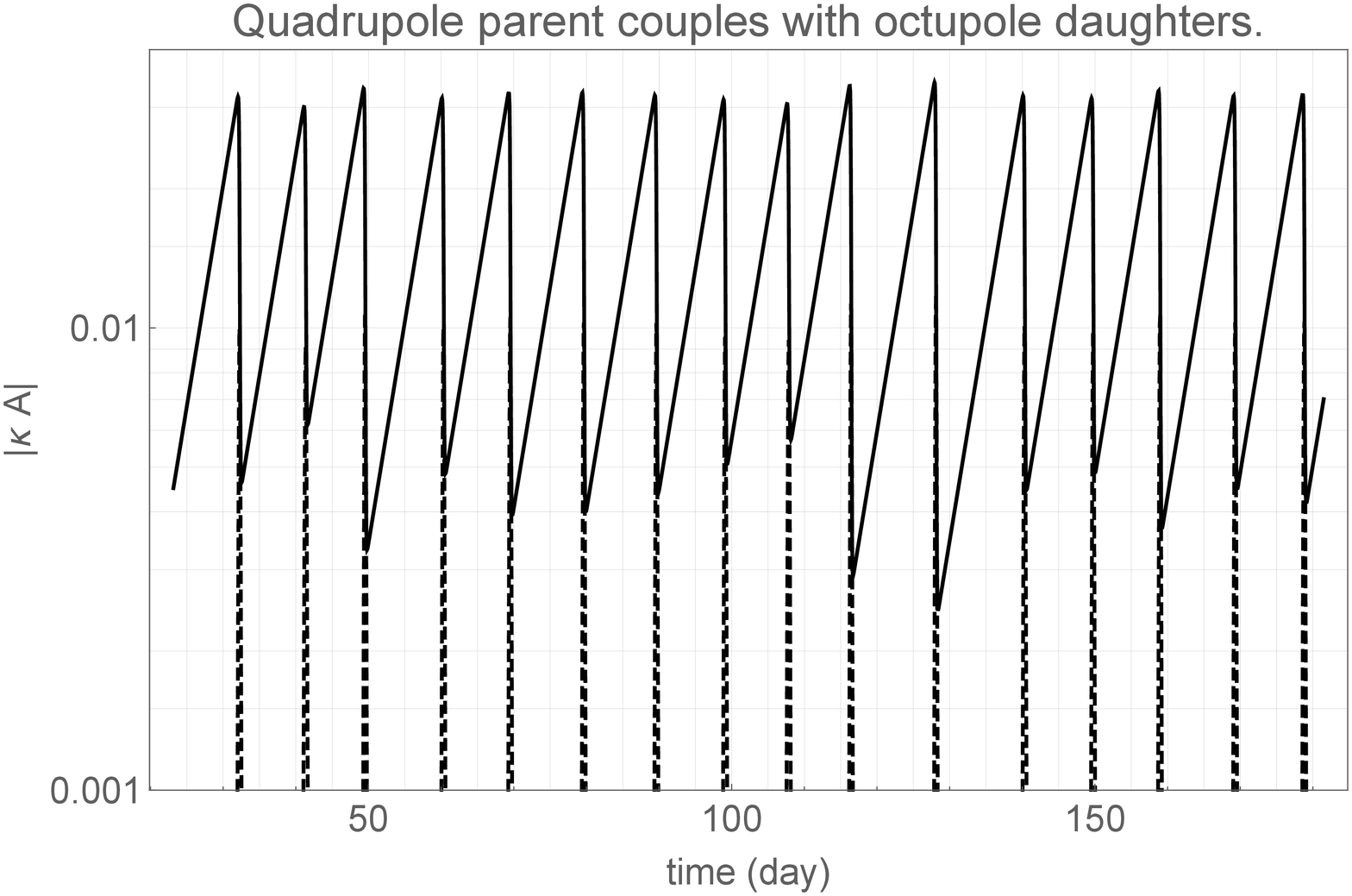}
\includegraphics[width=0.99\linewidth]{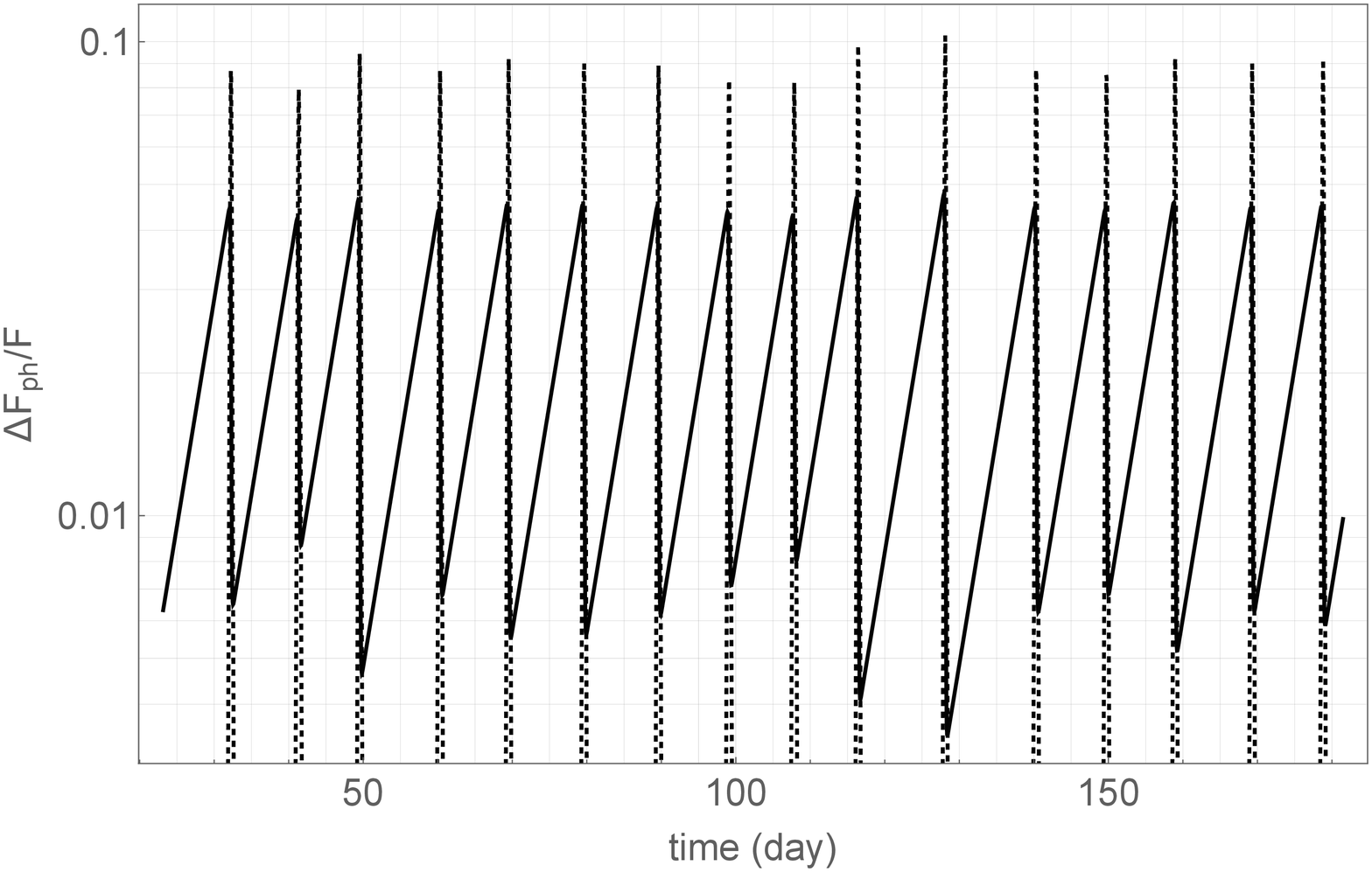}
\caption{\label{fig:outburst-23} Limit cycles involving a quadrupole parent mode coupled to identical octupole daughter modes with oscillation periods $P_p\approx 12.1145\,\mathrm{min}$, and $P_{d_1}=P_{d_2}\approx 24.2999\,\mathrm{min}$. Radial orders are $n_p=28$ and $n_{d_1}=n_{d_2}=82$. Rates of driving and damping are $\gamma_p\approx 1.1082\times 10^{-5}\,\mathrm{s^{-1}}$ and $\gamma_{d_1}=\gamma_{d_2}\approx -3.3758\times 10^{-4}\,\mathrm{s^{-1}}$. The frequency mismatch, $\delta\omega\equiv \omega_p-\omega_{d_1}-\omega_{d_2} \approx 2.5239\times 10^{-5} \,\mathrm{rad\, s^{-1}}$. The modes are calculated based on the same WD model as indicated in the caption of Figure~\refnew{fig:outburst-22}. The upper panel plots the dimensionless $|\kappa A|$ for the parent mode (solid line) and the first daughter mode (dashed line). The lower panel plots their photometric fluctuations, assuming the peak $|\kappa A_p|\approx 0.04$ corresponds to $10^{33}\,\mathrm{erg}$ for the parent mode.}
\end{figure}

Integrating Equations (1), (2) and (3) in \cite{Wu-Goldreich-IV},\footnote{We filter out the rapidly oscillating part of each modal amplitude by substituting a modified amplitude $A_j\exp(-i\omega_j t)$ ($j=p\, d_1\, d_2$) in these equations.} we present in Figures~\refnew{fig:outburst-22} and \refnew{fig:outburst-23} two examples of limit cycles involving the same $\ell=2$ parent mode coupled to different pairs of daughter modes.  In the former, the daughters are also $\ell=2$ and the frequency mismatch, $\delta\omega\ll \gamma_d$ is even smaller than $\gamma_p$.  In the latter, the daughters are identical twins with $\ell=3$ and although $\delta\omega<\gamma_d$, it is substantially larger than $\gamma_p$. In each example, recurrence times are in excess of 10 days. The limit cycles with two quadrupole daughter modes appear more irregular, a characteristic we observed in other cases with $|\delta\omega|<\gamma_p$. On the other hand, the limit cycle with two octupole daughter modes has $|\delta\omega|>\gamma_p$, and although not periodic, is more regular.\footnote{\cite{Wersinger} explored analogous behavior in a simple model of 3-mode coupling.}  

\section{Conclusion and Discussion}\label{sec:conclusion}
Our paper clarifies the physics behind the red edge of the DAV instability strip.  It also provides an
explanation for the photometric outbursts recently discovered from cool DAVs.

Regarding the red edge, we recapitulate the necessary and sufficient condition for overstable modes, $\omega>\max(\tau_c^{-1},\, \Lambb)$. Modes with angular frequencies below $\Lambb$ suffer enhanced radiative damping that overwhelms convective driving and are damped.  
Below the red edge, all overstable modes have $\omega>\Lambb>1$ which renders them invisible to photometric surveys.  This completes the conceptual understanding of the observational red edge. High-quality spectroscopic observations of bright DAs with $\Teff$ well below the red edge might reveal overstable modes through their surface velocities. This would be a valuable test for our explanation of the red edge.

We demonstrate that limit cycles due to sufficiently resonant 3-mode couplings quantitatively account for the energies, durations and repetition times of the outbursts observed by {\it Kepler}. Outbursting DAVs cluster close to the red edge because the condition for triggering limit cycles is preferentially satisfied there.

\section*{Acknowledgements}
We thank Yanqin Wu for her generous help, especially providing her programs and models for comparison with our work. We thank JJ Hermes, Chris Clemens, and Keaton Bell for comments on interpreting observations. Jing Luan is supported by the Theoretical Astronomy Center and Center for Integrative Planetary Science at University of California at Berkeley. 

\bibliography{Red-edge-paper}

\begin{thebibliography}{}
\expandafter\ifx\csname natexlab\endcsname\relax\def\natexlab#1{#1}\fi

\bibitem[{{Arras} {et~al.}(2003){Arras}, {Flanagan}, {Morsink}, {Schenk},
  {Teukolsky}, \& {Wasserman}}]{Arras}
{Arras}, P., {Flanagan}, E.~E., {Morsink}, S.~M., {et~al.} 2003, \apj, 591,
  1129

\bibitem[{{Bell} {et~al.}(2015){Bell}, {Hermes}, {Bischoff-Kim}, {Moorhead},
  {Montgomery}, {{\O}stensen}, {Castanheira}, \& {Winget}}]{Bell-2015}
{Bell}, K.~J., {Hermes}, J.~J., {Bischoff-Kim}, A., {et~al.} 2015, \apj, 809,
  14

\bibitem[{{Bell} {et~al.}(2016{\natexlab{a}}){Bell}, {Hermes}, {Montgomery},
  {Winget}, {Gentile Fusillo}, {Raddi}, \& {G{\"a}nsicke}}]{Bell-2016a}
{Bell}, K.~J., {Hermes}, J.~J., {Montgomery}, M.~H., {et~al.}
  2016{\natexlab{a}}, ArXiv e-prints, arXiv:1609.09097

\bibitem[{{Bell} {et~al.}(2016{\natexlab{b}}){Bell}, {Hermes}, {Montgomery},
  {Gentile Fusillo}, {Raddi}, {G{\"a}nsicke}, {Winget}, {Dennihy}, {Gianninas},
  {Tremblay}, {Chote}, \& {Winget}}]{Bell-2016b}
---. 2016{\natexlab{b}}, \apj, 829, 82

\bibitem[{{Brickhill}(1983)}]{Brickhill-1983}
{Brickhill}, A.~J. 1983, \mnras, 204, 537

\bibitem[{{Brickhill}(1990)}]{Brickhill-1990}
---. 1990, \mnras, 246, 510

\bibitem[{{Brickhill}(1991{\natexlab{a}})}]{Brickhill-1991a}
---. 1991{\natexlab{a}}, \mnras, 251, 673

\bibitem[{{Brickhill}(1991{\natexlab{b}})}]{Brickhill-1991b}
---. 1991{\natexlab{b}}, \mnras, 252, 334

\bibitem[{{Brink} {et~al.}(2005){Brink}, {Teukolsky}, \& {Wasserman}}]{Brink}
{Brink}, J., {Teukolsky}, S.~A., \& {Wasserman}, I. 2005, \prd, 71, 064029

\bibitem[{{Clemens}(1993)}]{Clemens-1993}
{Clemens}, J.~C. 1993, Baltic Astronomy, 2, 407

\bibitem[{{Cox}(1983)}]{Cox}
{Cox}, J.~P. 1983, {Theory of stellar pulsations.} ({Princeton University
  Press})

\bibitem[{{Cox} \& {Giuli}(1968)}]{Cox1}
{Cox}, J.~P., \& {Giuli}, R.~T. 1968, {Principles of stellar structure } ({New
  York: Gordon and Breach})

\bibitem[{{Dolez} \& {Vauclair}(1981)}]{Dolez-Vauclair}
{Dolez}, N., \& {Vauclair}, G. 1981, \aap, 102, 375

\bibitem[{{Dziembowski}(1982)}]{Dziembowski}
{Dziembowski}, W. 1982, \actaa, 32, 147

\bibitem[{{Goldreich} \& {Wu}(1999)}]{Goldreich-Wu-I}
{Goldreich}, P., \& {Wu}, Y. 1999, \apj, 511, 904

\bibitem[{{Hermes} {et~al.}(2015){Hermes}, {Montgomery}, {Bell}, {Chote},
  {G{\"a}nsicke}, {Kawaler}, {Clemens}, {Dunlap}, {Winget}, \&
  {Armstrong}}]{Hermes}
{Hermes}, J.~J., {Montgomery}, M.~H., {Bell}, K.~J., {et~al.} 2015, \apjl, 810,
  L5

\bibitem[{{Hermes} {et~al.}(2017){Hermes}, {G{\"a}nsicke}, {Kawaler}, {Greiss},
  {Tremblay}, {Gentile Fusillo}, {Raddi}, {Fanale}, {Bell}, {Dennihy}, {Fuchs},
  {Dunlap}, {Clemens}, {Montgomery}, {Winget}, {Chote}, {Marsh}, \&
  {Redfield}}]{Hermes-2017}
{Hermes}, J.~J., {G{\"a}nsicke}, B.~T., {Kawaler}, S.~D., {et~al.} 2017, \apjs,
  232, 23

\bibitem[{{Kumar} \& {Goodman}(1996)}]{Kumar-Goodman}
{Kumar}, P., \& {Goodman}, J. 1996, \apj, 466, 946

\bibitem[{{Mukadam} {et~al.}(2006){Mukadam}, {Montgomery}, {Winget}, {Kepler},
  \& {Clemens}}]{Mukadam}
{Mukadam}, A.~S., {Montgomery}, M.~H., {Winget}, D.~E., {Kepler}, S.~O., \&
  {Clemens}, J.~C. 2006, \apj, 640, 956

\bibitem[{{Paxton} {et~al.}(2011){Paxton}, {Bildsten}, {Dotter}, {Herwig},
  {Lesaffre}, \& {Timmes}}]{Paxton}
{Paxton}, B., {Bildsten}, L., {Dotter}, A., {et~al.} 2011, \apjs, 192, 3

\bibitem[{{Pesnell}(1987)}]{Pesnell}
{Pesnell}, W.~D. 1987, \apj, 314, 598

\bibitem[{{Townsend} \& {Teitler}(2013)}]{Townsend}
{Townsend}, R.~H.~D., \& {Teitler}, S.~A. 2013, \mnras, 435, 3406

\bibitem[{{Tremblay} {et~al.}(2015){Tremblay}, {Ludwig}, {Freytag}, {Fontaine},
  {Steffen}, \& {Brassard}}]{Tremblay-MLT}
{Tremblay}, P.-E., {Ludwig}, H.-G., {Freytag}, B., {et~al.} 2015, \apj, 799,
  142

\bibitem[{{van Kerkwijk} {et~al.}(2000){van Kerkwijk}, {Clemens}, \&
  {Wu}}]{van-Kerkwijk}
{van Kerkwijk}, M.~H., {Clemens}, J.~C., \& {Wu}, Y. 2000, \mnras, 314, 209

\bibitem[{{Wersinger} {et~al.}(1980){Wersinger}, {Finn}, \& {Ott}}]{Wersinger}
{Wersinger}, J., {Finn}, J.~M., \& {Ott}, E. 1980, The Physics of Fluids, 23,
  1142

\bibitem[{{Wu} \& {Goldreich}(1999)}]{Wu-Goldreich-II}
{Wu}, Y., \& {Goldreich}, P. 1999, \apj, 519, 783

\bibitem[{{Wu} \& {Goldreich}(2001)}]{Wu-Goldreich-IV}
---. 2001, \apj, 546, 469

\end{thebibliography}
\appendix

\section{Effective thermal timescale at the bottom of the surface CVZ}\label{sec:derivation}

We motivate why $\tau_c\gg \tau_\mathrm{b}$ as follows. Flux perturbations associated with a mode alternate sign on timescale $\pi/\omega$, much longer than both the sound crossing time, $H_p/c_s$, and the convective mixing time, $H_p/v_\mathrm{cv}$. Therefore, the CVZ essentially remains hydrostatic and, except for its thin superadiabatic layer, also isentropic. The total enthalpy of the CVZ increases by
\begin{eqnarray}
\Delta Q &=& \int_{\mathrm{CVZ}} dm\, c_p \delta T\approx 4\pi R^2\int_{z_\mathrm{ph}}^{z_b} dz\, \rho T \left(\partial s\over \partial T\right)_p \left(k_B\over m_p\right)\delta T\nonumber\\
&\approx &4\pi R^2 \int_{z_\mathrm{ph}}^{z_b} dz\, \rho {k_B\over m_p} T \left.\delta s\right|_p\sim 4\pi R^2 \Delta s_b \int_{z_\mathrm{ph}}^{z_b}dz\, \rho {k_B\over m_p} T \nonumber\\
&\sim& 4\pi R^2 \Delta s_b F \tau_\mathrm{b}\, ,
\end{eqnarray}
where we approximate $\delta s(z)\sim \Delta s_b$.  It takes the photosphere $\tau_c$ to get rid of this amount of heat, i.e.,
\begin{equation}
\tau_c= {\Delta Q\over 4\pi R^2\Delta  F_\mathrm{ph}}\, .
\end{equation}

There are two reasons why $\tau_c\gg\tau_b$. 
\begin{itemize}
\item First, the photosphere is an inefficient radiator. Because hydrogen is partially ionized, it must absorb a lot of heat to raise the temperature by a small amount;  most of the absorbed heat goes to ionize and/or excite hydrogen atoms. Moreover, given that $\Delta F_\mathrm{ph}=4 F (\Delta T_\mathrm{eff}/T_\mathrm{eff})$, the entropy at the photosphere must rise a lot to effect a small fractional increase of the photospheric flux.  \cite{Goldreich-Wu-I} estimate
\begin{equation}\label{eq:Dsph-DFph}
\Delta s_\mathrm{ph} \approx B {\Delta F_\mathrm{ph}\over F} \sim 10 {\Delta F_\mathrm{ph}\over F}\, ,
\end{equation}
in their Section 3.3. Our WD model with $T_\mathrm{eff}\approx 10860\,\mathrm{K}$ and $g\approx 10^8\,\mathrm{cm\, s^{-2}}$ also has $B\sim 10$.

\item Second, convection struggles to transport flux in the layer just below the photosphere. Because $F_\mathrm{cv}\sim \rho v_\mathrm{cv}^3$, $v_\mathrm{cv}$ approaches $c_s$ as the density drops. It follows from the mixing length prescription, $v_\mathrm{cv}^2\sim -(\alpha H_p)^2 g \rho_s (ds/dz)$, that almost all the entropy drop across the CVZ is concentrated in this thin layer, i.e. 
\begin{equation}
s_b-s_\mathrm{ph} \sim \left[ {ds\over dz} H_p\right]_\mathrm{ph}\, .
\end{equation}
This relation together with the mixing length prescription yields
\begin{equation}\label{eq:Dsb-Dsph-DFph}
\Delta s_b -\Delta s_\mathrm{ph} \sim C(s_b-s_\mathrm{ph}) {\Delta F_\mathrm{ph}\over F}\sim 5 {\Delta F_\mathrm{ph}\over F}\, ,
\end{equation}
as derived in Section 3.2 of \cite{Goldreich-Wu-I}. Our WD model at $T_\mathrm{eff}\approx 10860\,\mathrm{K}$ and $g\approx 10^8\,\mathrm{cm\, s^{-2}}$ has $C\sim 1$ and $(s_b-s_\mathrm{ph})\sim 5$.\footnote{In \cite{Goldreich-Wu-I}, equation (39) follows from equations (24) and (25), and its right hand side should be $ C(s_b-s_{\mathrm{ph}})\,\Delta F_{\mathrm{ph}}/F$. Thus $C$ in their equations (43) and (44) should be replaced by $C(s_b-s_{\mathrm{ph}})$.}

Combining Equations~\refnew{eq:Dsph-DFph} and \refnew{eq:Dsb-Dsph-DFph} and eliminating $\Delta s_\mathrm{ph}$, we obtain
\begin{equation}\label{eq:Dsb-DFph}
\Delta s_b =[B+C(s_b-s_\mathrm{ph})]{\Delta F_\mathrm{ph}\over F}\sim 15 {\Delta F_\mathrm{ph}\over F}\, .
\end{equation}
It indicates that the CVZ entropy has to rise a lot to produce a small fractional increase in the photospheric flux. 

\end{itemize}

\end{document}